\documentstyle[12pt]{article}
\newcommand{\pslash}{\mbox{$\not \!p$}}
\newcommand{\qslash}{\mbox{$\not \!q$}}

\begin{document}
\title{On the pattern of asymmetries in the pole model of
weak radiative hyperon decays }
\author{
{P. \.{Z}enczykowski}$^*$\\
\\
{\it Department of Theoretical Physics,} \\
{\it Institute of Nuclear Physics,}\\
{\it Radzikowskiego 152,
31-342 Krak\'ow, Poland}\\
}
\maketitle
\begin{abstract}
We study the question whether
the pole-model VMD approach to weak radiative hyperon decays
can be made consistent with Hara's theorem and
still yield the pattern of asymmetries characteristic of the quark model.
It is found that an essential ingredient which governs the
pattern of asymmetries is the assumed off-shell behaviour
of the parity-conserving $1/2^--1/2^+-\gamma $ amplitudes.
It appears that this behaviour can be chosen in such a way that
the pattern  
characteristic of the quark model is obtained, 
and yet Hara's theorem satisfied.
As a byproduct, however,
all parity-violating amplitudes in weak radiative and nonleptonic hyperon
decays must then vanish in the $SU(3)$ limit.
This is in conflict with the observed size of weak meson-nucleon couplings. 

\end{abstract}
$^*$ Email address:zenczyko@solaris.ifj.edu.pl\\
\vskip 0.8in
\begin{center}REPORT \# 1790/PH, INP-Krak\'ow \end{center}
\newpage

\section{Introduction}
Weak radiative hyperon decays (WRHD's) present
a challenge to our theoretical
understanding.  Despite many years of theoretical studies, a
satisfactory description of these processes is still lacking.
For a review see ref.\cite{LZ} where current
theoretical and experimental situation in the field is presented.

The puzzle posed by WRHD's manifests itself as
a possible conflict between Hara's theorem \cite{Hara} and experiment.
Hara's theorem is 
formulated in the language of local field theory 
at hadron level, and is based on CP- and gauge-
invariance. 
It states that the parity-violating amplitude of the
$\Sigma ^+ \rightarrow p \gamma $ decay should vanish in the limit of SU(3)
flavour symmetry. 
For expected weak breaking of SU(3) symmetry
the parity-violating amplitude in question and, consequently, 
the $\Sigma ^+ \rightarrow p \gamma $ decay asymmetry should be small. 
Experiment \cite{Foucher} shows, however, that the asymmetry is large:
$\alpha (\Sigma ^+ \rightarrow p \gamma )= -0.72 \pm 0.086 \pm 0.045$.
Explanation of such a large value of this asymmetry
is even more difficult
when one demands a successful simultaneous description of the
experimental values of the asymmetries of three related WRHD's, namely
$\Lambda \rightarrow n \gamma $, $\Xi ^0 \rightarrow \Lambda \gamma $, 
and $\Xi ^0 \rightarrow \Sigma ^0 \gamma $.

Theoretical calculations may be divided into those 
performed totally at quark level (eg. \cite{KR,VS})
and those ultimately carried out at hadron level (eg. \cite{Gavela,Zen}).
Hadron-level calculations are based on the pole model, with 
Hara's theorem usually satisfied by construction.
The only exception is the hadron-level 
vector-meson dominance (VMD) symmetry approach of ref.\cite{Zen}
which admits a pole-model interpretation and yet violates the theorem. 
On the other hand, quark model calculation of
ref.\cite{KR} (and its phenomenological applications \cite{VS}), 
in spite of being explicitly CP- and gauge- invariant, directly
violate the theorem.
The problem is further confounded by the fact that experiment
seems to agree with the predictions of the quark (or VMD) model, and not with
those of the pole model satisfying Hara's theorem.
Putting aside the approach of ref.\cite{Zen},
for known pole and quark models
there exists an important difference between their predictions concerning
the pattern of the signs of asymmetries in the four WRHD's mentioned above.
For the set of asymmetries
($\Sigma ^+ \rightarrow p \gamma $, 
$\Lambda \rightarrow n \gamma $, $\Xi ^0 \rightarrow \Lambda \gamma $, 
$\Xi ^0 \rightarrow \Sigma ^0 \gamma $)
the pole model \cite{Gavela} predicts the pattern $(-,-,-,-)$,
while the quark model \cite{VS,LZ} gives $(-,+,+,-)$.
Experiment (and in particular the sign of 
the $\Xi ^0 \rightarrow \Lambda \gamma $
asymmetry \cite{James}) hints \cite{LZ} that it is the latter alternative that
is realized in Nature.
Apart from the quark model, there are two other approaches that yield 
the pattern  $(-,+,+,-)$.
The first one 
is the hadron-level $SU(6)_W \times$ VMD approach of ref.\cite{Zen}
which so far gives the best description of data \cite{LZ}.
The other is a diquark approach of ref.\cite{Zen98}.

The VMD prescription seems to violate Hara's theorem as well.
Although a connection between the quark model and VMD result has been proposed
\cite{Zen}, closer inspection \cite{Acta}
reveals that the origin of the violation
of Hara's theorem is slightly different in the two models.
In the quark model, the violation of Hara's theorem arises from 
bremsstrahlung diagrams in which photon is emitted from one
of the pair of quarks exchanging the $W$-boson.
The violation is connected with the
intermediate quark entering its mass-shell in the $q_{\gamma} \rightarrow 0$
limit.
The $SU(6)_W\times $VMD approach 
(related by symmetry to the standard pole model of nonleptonic hyperon decays)
admits a pole-model interpretation.
Then, the intermediate state is an excited $1/2^-$ state
which is not degenerate with external ground state baryon.
Hence, the intermediate excited baryon state cannot be on its mass shell.

The diquark approach \cite{Zen98} 
contains a few free parameters, among them the masses
of spin 0 and spin 1 diquarks.  In the limit when these masses are equal to
each other the approach yields the pattern $(-,+,+,-)$. 
Furthermore, all parity violating amplitudes are then
proportional to the $m_s-m_d$ mass difference and, consequently,
Hara's theorem is satisfied.
The pattern $(-,+,+,-)$ for the diquark approach 
looks a little bit like an accident since it holds
only when spin~0 - spin~1 symmetry is satisfied.
Still, the result of ref.\cite{Zen98} poses the question if one can find
other models which satisfy Hara's theorem and yet give the pattern $(-,+,+,-)$.

Specifically, the question that we put forward in this paper is:
can the phenomenological success of VMD \cite{LZ} 
be consistent with Hara's theorem?
We will show that the answer to this question is "yes".
However, consistency of the  phenomenological success 
of the $SU(6)_W \times $ VMD approach
 with Hara's theorem implies
that {\it dominant parts of \underline{all} 
parity violating WRHD's amplitudes
(as well as those of nonleptonic hyperon decays)
must vanish in the SU(3) limit}.
This markedly differs from the way in which Hara's theorem is satisfied in 
the standard pole model of ref.\cite{Gavela}.
That is, in ref.\cite{Gavela} it is only the $\Sigma ^+ \rightarrow p \gamma $
parity violating amplitude that vanishes in the SU(3) limit, while
three remaining 
relevant WRHD parity-violating amplitudes remain constant and nonzero.
The difference between the pole model of ref.\cite{Gavela} and the pole
model considered in this paper is connected to the
off-shell behaviour of the $B^* B \gamma $ couplings.
Throughout this paper, all our formulas will be consistent with Hara's
theorem: we will not refer to ref.\cite{KR} otherwise than in a discussion.

\section{Photon-baryon couplings}

Let us consider parity-violating, CP-conserving interaction of a photon
with spin $1/2^+$ baryons.
The most general 
conserved electromagnetic axial current of spin $1/2^+$ baryons 
may be written in this case as:
\begin{equation}
\label{eq:j5standard}
j_5^{\mu } = 
g_{1,kl}(q^2)
\overline{\psi}_k(q^2\gamma ^{\mu }-q^{\mu}\qslash )\gamma _5 \psi _l+
g_{2,kl}(q^2)
\overline{\psi}_k i\sigma ^{\mu \nu }\gamma _5 q_{\nu } \psi _l
\end{equation}
where $q=p_l-p_k$ and we use conventions of ref.\cite{BD} for $\gamma $ 
matrices.
Note the factor of $q^2$ in the first term of Eq.(\ref{eq:j5standard}).
Indices $l,k$ label initial and final baryons and may be different
(eg. $(l,k)=(\Sigma ^+,p)$, etc.).
Hermiticity and CP invariance of $j_5\cdot A$ coupling
require functions $g_1$, $g_2$ to be real
(see eg. ref.\cite{Okun}).
Furthermore, $g_1$ is symmetric and $g_2$ antisymmetric in baryon indices:
\begin{eqnarray}
\label{eq:symprop}
g_{1,kl}&=&g_{1,lk}\nonumber \\
g_{2,kl}&=&-g_{2,lk}
\end{eqnarray}
For real photons ($q^2=0$, $q\cdot A=0$) the coupling to a photon
of the first term in 
Eq.(\ref{eq:j5standard}) vanishes.
Thus, the only contribution may come from the second term.
Hara's theorem \cite{Hara} states that in the SU(3) limit
the function $g_{2,\Sigma ^+ p}$ must vanish. The reason is simple: in the
SU(3) limit wave functions of $\Sigma ^+$ and $p$ must be identical since
they are obtained from each other by a simple replacement $s \leftrightarrow d$.
Furthermore, photon is a U-spin singlet. 
Thus, function $g_{2,\Sigma ^+p}$ must be proportional to $g_{2,pp}$ 
(apart from the Cabibbo factor, nothing
changes when we replace $s$ by $d$ in $\Sigma ^+$).
Because of its antisymmetry the function $g_{2,pp}$ is, however, zero.
This proof does not specify, however, 
in what way the function $g_{2,\Sigma ^+ p}$
vanishes.  Furthermore, it says nothing about functions $g_{2,kl}$ for 
the remaining three WRHD's:
$\Lambda \rightarrow n \gamma $, $\Xi ^0 \rightarrow \Lambda \gamma $, 
and $\Xi ^0 \rightarrow \Sigma ^0 \gamma $.

In the pole model of ref.\cite{Gavela} WRHD's proceed in two stages:
a virtual decay of
the initial ground-state baryon $B_i$ into a photon and an
excited spin $1/2^-$ $B^*$ baryon followed by a  weak interaction 
transforming the latter
into a final ground-state baryon $B_f$
(a reverse order of interactions is of course also taken into account).
To describe these processes one has to know in particular the $B^*B\gamma $
couplings.

In ref.\cite{Gavela} these couplings 
are given in the form of a parity-conserving interaction of 
the photon with a current 
whose form (after setting $q^2=q\cdot A =0$)
is fully analogous to Eq.(\ref{eq:j5standard}):
\begin{equation}
\label{eq:jpcGavela}
j_{(2)}^{\mu}(B^*B) = f_{2,kl}(q^2) 
\overline{\psi}_ki\sigma ^{\mu \nu }\gamma _5 q_{\nu } \psi _l
\end{equation}
where a pair of indices $k$, $l$ denotes a pair of baryons $B$, $B^*$ 
under consideration, ie.
($k$,$l$) $\equiv $ ($B^*_k$, $B_l$) or ($B_k$, $B^*_l$).
Following ref.\cite{Okun} one can check that
hermiticity and CP invariance of $j_{(2)} \cdot A$ coupling
require function $f_2$ to be purely
imaginary and symmetric:
\begin{equation}
\label{eq:f2}
f_{2,kl}=f_{2,lk}
\end{equation}
In ref.\cite{Gavela} the corresponding function is stated to be
real and antisymmetric. This difference is inessential because one
can always absorb our purely imaginary phase of $f_2$ into the definition of
the spinor of the intermediate excited state. The relation valid for
both our convention and that of ref.\cite{Gavela} is
$f_2^{\dagger} = -f_2$.

There is one problem with Eq.(\ref{eq:jpcGavela}) that was not discussed
in ref.\cite{Gavela} at all: the form of the right-hand side of
Eq.(\ref{eq:jpcGavela}) is 
\underline{not} the most general form for the situation under consideration.
In fact, Eq.(\ref{eq:jpcGavela}) is fully correct only
when particles $B^*$, $B$ are {\it on their mass shells}. In the pole model,
however,
the intermediate excited states are certainly not on their mass shells.
Thus, the use of Eq.(\ref{eq:jpcGavela}) is not fully justified.

To substantiate our claim we shall consider the current:
\begin{equation}
\label{eq:j1pc}
j^{\mu }_{(1)}(B^*B)=f_{1,kl}(q^2)(-i )(p_k+p_l)_{\lambda }q_{\nu }
\epsilon ^{\lambda \mu \nu \rho}\overline{\psi}_k\gamma _{\rho }\psi _l
\end{equation}
which is quadratic in external momenta.
As before, $(k,l)$ =$(B^*_k,B_l)$ or $(B_k,B^*_l)$.
Hermiticity and CP invariance of the coupling of $j_{(1)}$ to a photon
require $f_1$ to be purely imaginary
and antisymmetric ($f_1^{\dagger}=f_1$ in phase-convention-independent form).
We observe that a form totally analogous to Eq.(\ref{eq:j1pc})
might also be used as an axial current relevant for describing 
the parity violating coupling of a photon to ground-state
baryons:
\begin{equation}
\label{eq:j5tilde}
\widetilde{j_5^{\mu }}=\tilde{g}_{kl}(q^2)(-i)(p_k+p_l)_{\lambda }q_{\nu }
\epsilon ^{\lambda \mu \nu \rho}\overline{\psi}_k\gamma _{\rho }\psi _l
\end{equation}
with initial and final spin $1/2^+$ baryons $k,l$.
Hermiticity and CP invariance of $\widetilde{j_5}\cdot A$ interaction 
require $\tilde{g}$ to be real and
symmetric:
\begin{equation}
\label{eq:gtilde}
\tilde{g}_{kl}=\tilde{g}_{lk}
\end{equation}
Using the identity
\begin{equation}
\label{eq:identity}
\gamma ^{\alpha }\gamma ^{\beta }\gamma ^{\mu } =
g^{\alpha \beta }\gamma ^{\mu }-g^{\alpha \mu }\gamma ^{\beta}
+g^{\beta \mu }\gamma ^{\alpha }
-i\gamma _5 \epsilon ^{\alpha \beta \mu \nu} \gamma _{\nu }
\end{equation}
it is straightforward to show 
that
\begin{eqnarray}
\label{eq:basicidentity}
-i (p_k+p_l)_{\lambda }q_{\nu }
\lefteqn{
\epsilon ^{\lambda \mu \nu \rho}\overline{\psi}_k\gamma _{\rho }\psi _l =}
\nonumber \\
&\overline{\psi}_k(q^2\gamma ^{\mu }-q^{\mu }\qslash)\gamma _5\psi _l+
\overline{\psi}_k(\pslash_k
i\sigma ^{\mu \nu }\gamma _5 q_{\nu }-
i\sigma ^{\mu \nu }\gamma _5 q_{\nu }\pslash_l)\psi _l&
\end{eqnarray}
Thus, for particles $k$, $l$ on their mass shell the 
current $\widetilde{j_5}$ of 
Eq.(\ref{eq:j5tilde}) reduces to the current $j_5$
of Eq.(\ref{eq:j5standard}) with
$g_{1,kl}=\tilde{g}_{kl}$ and $g_{2,kl}=(m_k-m_l)\tilde{g}_{kl}$.
Interaction with real transverse photons of the first term 
on the rhs of Eq.(\ref{eq:basicidentity}) vanishes.
As to the second term, 
please note that the obtained function $g_{2,kl}$ is antisymmetric and that it
vanishes
for equal masses of baryons $k$, $l$.
Although for the parity-conserving current $j^{\mu }_{(1)}(B^*B)$ 
the identity of 
Eq.(\ref{eq:basicidentity}) also holds, in the pole model
of WRHD's one cannot in general replace $\pslash_k$ and
$\pslash_l$ by the corresponding baryon masses: the intermediate baryons $B^*$
are not on their mass shell.  We shall see later what are the consequences
of this lack of sufficient generality of the current of Eq.(\ref{eq:jpcGavela}).

\section{Parity-violating amplitudes in pole model}
The pole model is built from two basic building blocks. 
The first describes weak interaction, the second  - electromagnetic
emission of a photon. Parity violation comes
from weak interactions which transform ground-state baryons into excited 
spin $1/2^-$ baryons and vice versa. 

The parity-violating weak transitions are described by 
\begin{equation}
\label{eq:weak}
a_{kl}\overline{\psi}_k\psi _l
\end{equation}
where the pair of indices $k$, $l$ describes a pair of baryons ($B$,$B^*$), ie.
$(k,l)$ $=$ $(B^*_k,B_l)$ or $(B_k,B^*_l)$. 
Hermiticity and CP invariance require $a$ to be purely imaginary
and antisymmetric:
\begin{equation}
\label{eq:akl}
a_{kl}=-a_{lk}
\end{equation}
(Again we differ in conventions with ref.\cite{Gavela} where $a$ is real
and symmetric. A convention-independent condition is $a^{\dagger}=a$.)

The electromagnetic emission is described by coupling the photon to the sum 
$j(B^*B)$ of currents of Eqs.(\ref{eq:jpcGavela},\ref{eq:j1pc}):
\begin{eqnarray}
\label{eq:gencurrent}
j^{\mu }(B^*B) & = & f_{1,kl}(q^2)(-i)(p_k+p_l)_{\lambda }q_{\nu }
\epsilon ^{\lambda \mu \nu \rho}\overline{\psi}_k\gamma _{\rho }\psi _l+
\nonumber \\
&& f_{2,kl}(q^2) 
\overline{\psi}_ki\sigma ^{\mu \nu }\gamma _5 q_{\nu } \psi _l
\end{eqnarray}

The calculation of ref.\cite{Gavela} corresponds to $f_1=0$, $f_2\ne 0$ and
leads to the pattern $(-,-,-,-)$ (see eg. ref.\cite{LZ}).
Since this case was studied elsewhere \cite{Gavela,LZ},
we will consider it only in a discussion, a little later.  
The really novel feature is
the first term ($f_1$) on the right hand side 
of Eq.(\ref{eq:gencurrent}).
We turn now to the evaluation of its effects.
We will show that this term  generates asymmetry
pattern $(-,+,+,-)$.

There are two pole-model diagrams (Fig.1a,b) contributing to the decay 
$B_i \rightarrow B_f \gamma $.
The amplitude corresponding to these diagrams is built from our basic blocks
in a simple way.
Weak interaction (symbolized by blobs in Fig.1) is described by 
Eq.(\ref{eq:weak}) while the electromagnetic current by 
Eq.(\ref{eq:gencurrent}). 
In addition, there must be a pole factor $1/(p^2-m_*^2)$ corresponding to the
propagation of the off-shell excited baryon $B^*$.

Using the first term ($j_{(1)}$) of the current of Eq.(\ref{eq:gencurrent})
the following expression corresponds then to Fig. 1a:
\begin{equation}
\label{eq:expFig1a}
f_{1,fk^*}(-i)(p_f+p_{k^*})_{\lambda}q_{\nu }\epsilon ^{\lambda \mu \nu \rho}
\overline{u}_f\gamma _{\rho}u_{k^*} \cdot \frac{1}{p^2_i-m^2_{k^*}} \cdot
a_{k^*i} \overline{u}_{k^*}u_i
\end{equation}
where $k^*$ labels intermediate excited states 
(summation over admissible $k^*$ is implied).
The contribution corresponding to Fig. 1b is
\begin{equation}
\label{eq:expFig1b}
a_{fk^*} \overline{u}_f u_{k^*} \cdot \frac{1}{p^2_f-m^2_{k^*}} \cdot
f_{1,k^*i}(-i)(p_{k^*}+p_i)_{\lambda }q_{\nu } 
\epsilon ^{\lambda \mu \nu \rho} \overline{u}_{k^*}\gamma _{\rho } u_i
\end{equation}
with appropriate $m_{k^*}$, different from that in Eq.(\ref{eq:expFig1a}).
However, since we are mainly concerned with the limit $m_s-m_d \rightarrow 0$,
for our purposes it is sufficient 
to consider $1/2^+$ $-$ $1/2^-$ mass splitting
to be much larger than $m_s-m_d$. Thus, 
we may put the same $m_{k^*}$ everywhere.
Upon summing the above two contributions and replacing the
factor $u_{k^*}\overline{u}_{k^*}$ by $\pslash_{k^*}+m_{k^*}$,
we act with $\pslash_{k^*}$ on $u_i$ ($\overline{u}_f$) for
the contributions of Fig 1a (1b) respectively. This yields $m_i$ ($m_f$).
Using $p_i^2=m^2_i$ and $p^2_f=m^2_f$ we obtain the total 
pole-model contribution from $f_1$ terms:
\begin{equation}
\label{eq:totalf1}
-i(p_i+p_f)_{\lambda }q_{\nu } \epsilon ^{\lambda \mu \nu \rho}
\overline{u}_f\gamma _{\rho } u_i \cdot \left \{
\frac{f_{1,fk^*}a_{k^*i}}{m_i-m_{k^*}}+
\frac{a_{fk^*}f_{1,k^*i}}{m_f-m_{k^*}}
\right \}
\end{equation} 
Now, for real photons and external baryons on their mass shell the factor
in front of the braces in Eq.(\ref{eq:totalf1}) can be reduced using 
Eq.(\ref{eq:basicidentity}). In this way,  Eq.(\ref{eq:totalf1}) is
brought into our final form and the parity-violating WRHD amplitude is
obtained from:
\begin{equation}
\label{eq:finalform}
(m_f-m_i)
\left \{
\frac{f_{1,fk^*}a_{k^*i}}{m_i-m_{k^*}}+
\frac{a_{fk^*}f_{1,k^*i}}{m_f-m_{k^*}}
\right \}
\cdot \overline{u}_fi\sigma ^{\mu \nu }q_{\nu } \gamma _5 u_i A_{\mu }
\end{equation}
As Eq.(\ref{eq:finalform}) shows, \underline{all} parity-violating
WRHD amplitudes vanish now in the limit $m_i \rightarrow m_f$.
Furthermore, this vanishing does not come about as a result of the
cancellation between the contributions from 
the $s$- and $u$- channel poles as in ref.\cite{Gavela}.
In fact, for $f=i$ the denominators of the two terms in braces are identical 
and the same can be shown to hold for the numerators since:
1) $f_{1,fk^*}=f_{1,ik^*}=-f_{1,k^*i}$ and
2) $a_{fk^*}=a_{ik^*}=-a_{k^*i}$ leads to
$f_{1,fk^*}a_{k^*i}= (-f_{1,k^*i})(-a_{fk^*})$.
One can also easily see that under $i \leftrightarrow f$
interchange the expression in braces in Eq.(\ref{eq:finalform})
is symmetric, ie. $\{\ldots \}_{if}=+\{\ldots \}_{fi}$,
and therefore the whole expression $(m_f-m_i)\{\ldots \}$ is 
antisymmetric , in agreement with the second of Eqs.(\ref{eq:symprop}).

Let us now try to use the current $j_{(1)}$ while putting 
intermediate baryons $B^*$
on their mass shell.
For real transverse photons
the current $j_{(1)}$ of Eq.(\ref{eq:j1pc}) may be then reexpressed using
the simplified version of Eq.(\ref{eq:basicidentity}):
\begin{equation}
\label{eq:j1pconshell}
-i (p_k+p_l)_{\lambda }q_{\nu }
\epsilon ^{\lambda \mu \nu \rho}\overline{u}_k\gamma _{\rho }u_l=
(m_k-m_l)\overline{u}_ki\sigma ^{\mu \nu}q_{\nu } \gamma _5 u_l 
\end{equation}
for ($k,l$) = ($B^*_k,B_l$) or 
($B_k,B^*_l$).
The electromagnetic currents in Eqs.(\ref{eq:expFig1a},\ref{eq:expFig1b}) 
are then replaced by
\begin{equation}
\label{eq:GavFig1a}
f_{1,k^*i}(m_{k^*}-m_i)
\overline{u}_{k^*}i\sigma ^{\mu \nu }q_{\nu }\gamma _5 u_i 
\end{equation}
for Fig. 1a (Eq.(\ref{eq:expFig1a}))
and
\begin{equation}
\label{eq:GavFig1b}
f_{1,fk^*}(m_f-m_{k^*})
\overline{u}_{f}i\sigma ^{\mu \nu }q_{\nu }\gamma _5 u_{k^*} 
\end{equation}
for Fig. 1b (Eq.(\ref{eq:expFig1b})).
Please note that now the factors $f_{1,kl^*}(m_k-m_{l*})$ 
multiplying spinorial expressions in 
Eqs.(\ref{eq:GavFig1a},\ref{eq:GavFig1b})
have symmetry properties of the $f_2$ factors, ie. they are
symmetric under $k\leftrightarrow l^*$ interchange, as in Eq.(\ref{eq:f2}).
We might write $\tilde{f}_{2,fk^*}\equiv f_{1,fk^*}(m_f-m_{k^*})$ and 
$\tilde{f}_{2,k^*i}\equiv f_{1,k^*i}(m_{k^*}-m_i)$ with $\tilde{f}_2$ symmetric,
and thus fully analogous to
$f_2$ in Eqs.(\ref{eq:jpcGavela},\ref{eq:gencurrent}).
Consequently, results of ref.\cite{Gavela} should follow. Indeed,
applying the procedure described above for the true current $j_{(1)}$ we
obtain now the counterpart of Eq.(\ref{eq:finalform}) for the current
of Eqs.(\ref{eq:GavFig1a},\ref{eq:GavFig1b}):
\begin{equation}
\label{eq:Gavform}
\left \{
\frac{f_{1,fk^*}(m_f-m_{k^*})a_{k^*i}}{m_i-m_{k^*}}+
\frac{a_{fk^*}f_{1,k^*i}(m_{k^*}-m_i)}{m_f-m_{k^*}}
\right \}
\cdot
\overline{u}_fi\sigma ^{\mu \nu }q_{\nu } \gamma _5 u_i
\end{equation}

Now, for $f=i$ the denominators of the two terms in Eq.(\ref{eq:Gavform})
are identical but the numerators differ in sign since:
$f_{1,fk^*}=f_{1,ik^*}=-f_{1,k^*i}$,
$a_{fk^*}=a_{ik^*}=-a_{k^*i}$ and $(m_f-m_{k^*})=-(m_{k^*}-m_i)$.
Thus, for $f=i$ the two terms in Eq.(\ref{eq:Gavform}) cancel.
This is precisely the case considered in ref.\cite{Gavela} where only
the current $j_{(2)}$ was considered and the
cancellation between the two diagrams of Fig.1 was invoked as a way in
which Hara's theorem is satisfied. In ref.\cite{Gavela}
such a cancellation does not occur, however, for the remaining three relevant 
WRHD's, namely
$\Lambda \rightarrow n \gamma $, $\Xi ^0 \rightarrow \Lambda \gamma $, 
and $\Xi ^0 \rightarrow \Sigma ^0 \gamma $.

\section{Discussion}

Phenomenologically, the most successful model seems to be the VMD model
of ref.\cite{Zen} (and its update in ref.\cite{LZ}). In the VMD approach 
 the crucial assumption (apart from the VMD
prescription) is the assumed $SU(6)_W$ symmetry relating WRHD's to the
well measured experimentally nonleptonic hyperon decays (NLHD's). Thus,
the size and the pattern of parity violating WRHD amplitudes are determined
by symmetry from NLHD's. 

The symmetry structure of the parity-violating WRHD and NLHD amplitudes of 
refs.\cite{Zen} may be understood in terms of the pole model.
In view of:

\noindent
(1) considerations of the preceding section in which two different 
possible patterns of WRHD asymmetries were obtained in the pole model, and

\noindent
(2) the symmetry 
connection between WRHD's and NLHD's that forms the basis of the 
successful approach of refs.\cite{Zen}

\noindent
it is pertinent to discuss nonleptonic hyperon decays in the pole model along
the lines of the preceding section and to study the relation between the
symmetry structures of WRHD's and NLHD's.  This is what we will turn to now.

For the sake of further discussion let us assume that masses of octet
pseudoscalar mesons are negligible, $m^2_P \approx 0$.
Thus, we shall discuss
the parity-violating CP-conserving
amplitudes for the $B_i \rightarrow P^0 B_f$ 
couplings with $P^0$ a $CP=-1$
pseudoscalar meson ($\pi^0$  or $\eta _8$)
and $B_{i,f}$ - ground-state baryons.
Consider
the following coupling:
\begin{equation}
\label{eq:pvNLHD}
b^{(0)}_{fi}\overline{u}_fu_iP^0+
b^{(1)}_{fi}\overline{u}_f\qslash u_iP^0+
b^{(2)}_{fi}\overline{u}_f(-i\sigma _{\mu \nu}(p_f+p_i)^{\mu}q^{\nu})u_iP^0
\end{equation}
where (by CP-invariance and hermiticity)
all $b^{(n)}$ are imaginary, with $b^{(0)}_{fi}$, $b^{(2)}_{fi}$
antisymmetric and $b^{(1)}_{fi}$ symmetric
under $i \leftrightarrow f$ interchange.
For baryons $B_f$, $B_i$ on mass shell
the coupling of Eq.(\ref{eq:pvNLHD}) may be rewritten ($q^2=m^2_P$) as
\begin{equation}
\label{eq:pvNLHDstandard}
\left\{b^{(0)}_{fi}+(m_i-m_f)b^{(1)}_{fi}+
[(m_i-m_f)^2-m^2_{P}]b^{(2)}_{fi}\right\}
\overline{u}_fu_iP^0
\end{equation}
where  we may put $m^2_{P } = 0$.
The a priori possible term
$b^{(1')}_{fi}\overline{u}_f(\not \!p_f+\not \!p_i)u_iP^0$, 
linear in external momenta,
may be absorbed into the $b^{(0)}_{fi}$ term.

In the pole model of NLHD's the couplings of Eq.(\ref{eq:pvNLHDstandard})
arise from the parity-violating weak
transition of Eq.(\ref{eq:weak}) followed by parity-conserving
$\pi ^0$ (or $\eta _8$) emission 
from the excited spin $1/2^-$ baryon (a reverse order of interactions
is also taken into account).
Consider parity-conserving $P^0$ emission couplings described by:
\begin{equation}
\label{eq:pcexcited}
f^{(0)}_{kl}\overline{u}_ku_lP^0+
f^{(1)}_{kl}\overline{u}_k\not \!qu_lP^0+
f^{(2)}_{kl}\overline{u}_k(-i\sigma _{\mu \nu}(p_k+p_l)^{\mu}q^{\nu})u_lP^0
\end{equation}
with $(k,l)=(B^*_k,B_l)$ or $(B_k,B^*_l)$.
Hermiticity and CP-invariance require all $f^{(n)}_{kl}$ to be real
with $f^{(0)}_{kl}$, $f^{(2)}_{kl}$ symmetric and $f^{(1)}_{kl}$
asymmetric under $k \leftrightarrow l$ interchange.
Since excited intermediate spin $1/2^-$ baryon is not on its mass shell
we are not allowed to replace Eq.(\ref{eq:pcexcited}) 
by a momenta-independent form analogous to Eq.(\ref{eq:pvNLHDstandard}).
(In Eq.(\ref{eq:pcexcited}) we have neglected an a priori possible term 
$f_{kl}^{(1')}\overline{u}_k(\pslash_k+\pslash_l)u_lP^0$;
calculation shows that its effect is fully analogous to that of 
the $f^{(0)}$ term.)
Working out the pole model contributions from various terms of
Eq.(\ref{eq:pcexcited}) we obtain (as in the previous section)

(1) from the $f^{(0)}$ term:
\begin{equation}
\label{eq:fromf0}
\left\{
\frac{f^{(0)}_{fk^*}a_{k^*i}}{m_i-m_{k^*}}+
\frac{a_{fk^*}f^{(0)}_{k^*i}}{m_f-m_{k^*}}
\right\}
\overline{u}_fu_iP^0
\end{equation}
with the factor in braces antisymmetric under $i \leftrightarrow f$ interchange
(this is the term usually 
considered in papers on nonleptonic hyperon decays),

(2) from the $f^{(1)}$ term
\begin{equation}
\label{eq:fromf1}
(m_i-m_f)\left \{
\frac{f^{(1)}_{fk^*}a_{k^*i}}{m_i-m_{k^*}}+
\frac{a_{fk^*}f^{(1)}_{k^*i}}{m_f-m_{k^*}}
\right \}
\overline{u}_fu_iP^0
\end{equation}
with the factor in braces  
symmetric under $i \leftrightarrow f$ interchange,

(3) from the $f^{(2)}$ term 
\begin{equation}
\label{eq:fromf2}
[(m_i-m_f)^2-m^2_{P}]\left \{
\frac{f^{(2)}_{fk^*}a_{k^*i}}{m_i-m_{k^*}}+
\frac{a_{fk^*}f^{(2)}_{k^*i}}{m_f-m_{k^*}}
\right \}
\overline{u}_fu_iP^0
\end{equation}
with the factor in braces antisymmetric under $i \leftrightarrow f$ interchange.
Thus, the pole model yields specific predictions for $b^{(0)}_{fi}$,
$b^{(1)}_{fi}$, and $b^{(2)}_{fi}$ of Eq.(\ref{eq:pvNLHDstandard}), 
which are given by factors in braces in
Eqs.(\ref{eq:fromf0},\ref{eq:fromf1},
\ref{eq:fromf2}).

Assuming now that one of the two patterns of parity-violating NLHD
amplitudes (corresponding to the symmetry or antisymmetry of the factor
in braces) is dominant, there appears the question
which pattern is actually realized in Nature.

Calculations of Desplanques, Donoghue and Holstein (ref.\cite{DDH}) 
and those of ref.\cite{Zen} correspond to the pattern obtained 
from terms $f^{(0)}$ or $f^{(2)}$, which coincides with the predictions
of current algebra. For the sake of comparison 
with Eqs.(\ref{eq:fromf0},\ref{eq:fromf1},\ref{eq:fromf2}) 
in Table I we give a few selected amplitudes corresponding 
to the symmetry pattern of these references.
Table I explicitly demonstrates
 the antisymmetry of the factor $\{\ldots\}$ under
$\Sigma ^+ \leftrightarrow p$  ($p \leftrightarrow p$) interchange
and the cancellation between the contributions from diagrams (1a) and (1b)
for $f=i$:
for $pp\pi ^0$ case antisymmetry ensures vanishing of the total contribution
to the parity-violating $pp\pi ^0 $ coupling.
This is also what current algebra gives \cite{DDH} since
$\langle p \pi ^0|H^-_W|p\rangle \propto \langle p|[I_3,H^+_W]|p\rangle = 0$.
Such vanishing occurs also for $\Sigma ^+ \rightarrow p U^0$ coupling
where $U^0 = (\sqrt{3} \pi ^0 +\eta _8)/2$, a U-spin singlet.
\\

Table I

\noindent
Contribution of diagrams (1a) and (1b) to selected parity-violating 
$BB'P^0$ amplitudes. 

\begin{tabular}{|l|c|c|}
\hline
&diagram (1a)&diagram(1b)
\\[0.2cm]
\hline
$\langle p \pi ^0|H^-_W|\Sigma ^+\rangle $
&$-\frac{1}{6\sqrt{2}}c$
&$\frac{1}{2\sqrt{2}}b$
\\[0.2cm]
$\langle \Sigma ^+ \pi ^0|H^-_W|p \rangle$
&$-\frac{1}{2\sqrt{2}}b$
&$\frac{1}{6\sqrt{2}}c$
\\[0.2cm]
\hline
$\langle p \pi ^0|H^-_W| p\rangle$
&$\left(-\frac{1}{2\sqrt{2}}b-\frac{1}{6\sqrt{2}}c\right) \cot \theta _C$
&$\left(\frac{1}{2\sqrt{2}}b+\frac{1}{6\sqrt{2}}c\right) \cot \theta _C$
\\[0.2cm]
$\langle p U^0|H^-_W|\Sigma ^+\rangle$
&$-\frac{1}{2\sqrt{6}}b-\frac{1}{6\sqrt{6}}c$
&$\frac{1}{2\sqrt{6}}b+\frac{1}{6\sqrt{6}}c$
\\[0.2cm]
\hline
\end{tabular}
\\
\\

In Table I, the $b$-term originates from $W$-exchange diagrams, 
while the $c$-term
represents hadronic loop/quark-sea contribution \cite{DGsea}.
Although $W$-exchange seems to contribute to diagram (1b) only,
this does not mean that individual contributions from $W$-exchange with
nonstrange intermediate excited baryons are all zero.  They do not vanish
but they all cancel among themselves (cf. \cite{LeYaouanc}).
Experimental data on NLHD's cannot determine which of the two patterns
(corresponding to $f^{(0)}/f^{(2)}$ or $f^{(1)}$) is correct.
This is so because in all 
$\pi ^0$ emission amplitudes the $b$-terms come solely from diagrams
(1b) and the $c$-terms - solely from diagrams (1a).
Since the size and sign of  $c$ is a phenomenological parameter it is
impossible to differentiate between the two patterns.
If $\eta _8$ ($U^0$) emission were kinematically allowed, 
this would be possible:
cancellation of two contributions to the
$\langle p U^0|H^-_W|\Sigma ^+\rangle$ amplitude would be replaced by 
constructive interference from diagrams (1a) and (1b).

Let us now go back to WRHD's.
The connection between NLHD's and WRHD's is achieved in ref.\cite{Zen} 
by considering the
combined flavour-spin symmetry
$SU(6)_W$.  This symmetry is suited for the description of two-body decays
because spin generators of $SU(2)_W$ commute with Lorentz boosts along
decay axis \cite{LipMesh65}.  
Consequently, if one wants to apply $SU(2)_W$ it is appropriate 
to choose one of Lorentz frames obtained
from the initial particle rest frame  by boosts along
decay axis. Thus, we choose any frame in which 
${\bf p}_i+{\bf p}_f = \lambda {\bf q}$ with arbitrary $\lambda $.

For further discussion let us recall the following identity:
\begin{equation}
\label{eq:sigmagammaident}
\overline{u}_fi\sigma ^{\mu \nu}\gamma _5q_{\nu }u_i =
(m_f-m_i)\overline{u}_f\gamma ^{\mu }\gamma _5u_i - 
(p_i+p_f)^{\mu }\overline{u}_f\gamma _5 u_i
\end{equation}

After fixing the gauge to be the Coulomb one 
($A_0=0$, ${\bf A}\cdot {\bf q}=0$),
the second term
on the right hand side of Eq.(\ref{eq:sigmagammaident}) decouples from 
the photon.
Thus, in the $SU(2)_W$-symmetric framework, the terms 
$\overline{u}_fi\sigma ^{k \nu}\gamma _5q_{\nu }u_iA_{k}$ and 
$\overline{u}_f\gamma ^{k }\gamma _5u_iA_{k}$ 
lead to amplitudes proportional
to each other, the coefficient of proportionality being $m_i-m_f$.
Consequently,  the model of ref.\cite{Gavela}
($f_1=0$, $f_2\ne 0$) generates the same amplitudes as
\begin{equation}
\label{eq:Gavgamma}
(m_i-m_f) B^{(1)}_{if}\cdot \overline{u}_f\gamma ^{k }\gamma _5u_iA_{k}
\end{equation}
where $B^{(1)}_{if}$ denotes the term (asymmetric under $i \leftrightarrow f$ 
interchange) in braces in Eq.(\ref{eq:Gavform}) with 
$\tilde{f}_2 $ replaced by $ f_2$.
For the present paper ($f_1\ne 0$, $f_2=0$), 
Eq.(\ref{eq:finalform}) corresponds to
\begin{equation}
\label{eq:thispaper}
(m_i-m_f)^2 B^{(2)}_{if} 
\cdot \overline{u}_f\gamma ^{k }\gamma _5u_iA_{k}
\end{equation}
where $B^{(2)}_{if}$ 
denotes the term
(symmetric under $i\leftrightarrow f$)
in braces in Eq.(\ref{eq:finalform}).

The result of Kamal-Riazuddin \cite{KR} corresponds to the expression
$(m_i-m_f)^0 B^{(0)}_{if} 
\cdot \overline{u}_f\gamma ^{k }\gamma _5u_iA_{k}$
with some symmetric $B^{(0)}_{if}$.

In general, the factors $B^{(k)}_{if}$ do not vanish for $m_i=m_f$. 
Symmetry properties of factors $B^{(0)}$ and $B^{(2)}$ are identical and,
consequently, they lead to the same pattern of asymmetries: $(-,+,+,-)$.
On the other hand, dominance of the $B^{(1)}$ term would 
lead to the pattern $(-,-,-,-)$.

If new experiments confirm the pattern $(-,+,+,-)$ which seems to be favoured
by the older data (refs. \cite{LZ}), it will mean that the dominant 
parts of \underline{all} parity violating 
WRHD amplitudes are proportional to an \underline{even} power of $m_i-m_f$.
Thus, one of two possibilities below must hold. Either

\noindent
(1) Hara's theorem is violated as in the quark model calculations of 
ref. \cite{KR} with $B^{(0)}_{if}\ne 0$, or

\noindent
(2) Hara's theorem is satisfied 
as a byproduct of vanishing 
(in the limit $m_i \rightarrow m_f$)
of \underline{all} parity-violating WRHD amplitudes.
(This vanishing may be approximate for those decays where a nonzero
$B^{(1)}$ of Eq.(\ref{eq:Gavgamma}) may contribute).
This corresponds to $B^{(0)}=0$, 
$(m_{\Sigma}-m_N)B^{(2)} \gg B^{(1)} \approx 0$. 
In this case,
the observed large asymmetry of $\Sigma ^+ \rightarrow p \gamma$ decay
should not surprise us too much.  To say that the size of the relevant
parity-violating amplitude is "large" means that we have to compare it
with some standard size. Thus, we should compare the 
 $\Sigma ^+ \rightarrow p \gamma$ amplitude with
other parity-violating amplitudes of WRHD's. However, since they all
vanish in the SU(3) limit in the same way as the $\Sigma ^+ \rightarrow p
\gamma $ amplitude does, 
the relative size of the latter amplitude is large indeed.

Within the $SU(6)_W \times $ VMD approach 
one expects that in NLHD's and WRHD's the terms of the same order
in $m_i-m_f$ are symmetry-related. Thus, if 
$SU(6)_W \times $ VMD predictions for the WRHD
asymmetries are borne out by the data 
and one insists that Hara's theorem is to be satisfied,
this would mean that
only the
contributions from $f^{(2)}$ terms should be present in NLHD's
and that, consequently, 
the parity-violating NLHD
amplitudes should vanish in the $SU(3)$ limit. 

However, since the mass
of the decaying particle is not a free parameter,
 one cannot differentiate between
contributions of type $f^{(0)}$ and $f^{(2)}$
using data on hyperon nonleptonic decays alone. Nonetheless,
instead of considering $\Delta S = 1$ decays, one may
study $\Delta S = 0$ parity-violating $NNM$ couplings, and try to see
if mass-dependence characteristic of $f^{(2)}$ (or perhaps $f^{(1)}$)
terms is present in these couplings.
In theoretical calculations 
mass-dependence characteristic of the $f^{(1)}$ term 
was obtained in the past \cite{Melosh}, leading to 
$A(n^0_-)$ of order $(m_n-m_p)/(m_{\Sigma }-m_p)$ 
$\approx 10^{-2}$ 
times the "best values" of
ref.\cite{DDH}.
If the NLHD and WRHD amplitudes are indeed proportional to 
$(m_i-m_f)^2$ 
as the signature $(-,+,+,-)$ and insistence on satisfying Hara's theorem would
demand, then
one would expect totally negligible weak parity-violating
$NN\pi $ and $NNV$ couplings.
At present, data seem to indicate that these couplings, although 
somewhat smaller than
the "best value" prediction of ref.\cite{DDH}, are nonetheless of the same
order \cite{AH85,DGH}.  Totally negligible value of weak
$NN\pi$ coupling is also possible \cite{Bouchiat91}.
However, the general order of magnitude of 
$NNM$ couplings is consistent
with the lack of the $(m_i-m_f)^2$ factor \cite{AH85,DGH}. 
Hence, although in principle it is possible that the signature $(-,+,+,-)$
for the WRHD asymmetries is consistent with Hara's theorem, the underlying
approach leads then to negligible weak $NNM$ couplings in disagreement
with experiment.

\section{Conclusions}
We have studied parity-violating WRHD amplitudes in the pole model.
In this model the properties of these amplitudes depend on the properties
of the parity-conserving $1/2^--1/2^+-\gamma $ couplings.
Two different conserved electromagnetic local baryonic currents have been
used for the description of the transition of an on-shell ground-state
baryon into an off-shell excited baryon (or vice versa).
Although the two currents become indistinguishable for a transition
between on-shell baryons, they are inequivalent when baryons are off-shell.
As a result, the two currents lead to different patterns of asymmetries
in weak radiative hyperon decays.
We have shown that in the pole model with Hara's theorem explicitly satisfied
it is still possible to obtain the asymmetry pattern  $(-,+,+,-)$ 
that is characteristic of the quark model.
Thus, the pattern $(-,+,+,-)$ is not an unmistakable sign of the violation
of Hara's theorem.
Phenomenological success of the $SU(6)_W \times$ VMD approach to WRHD's
may be understood as being consistent with Hara's theorem if the dominant parts
of \underline{all} WRHD and NLHD parity-violating amplitudes vanish
in the $SU(3)$ limit.
Although the success of the $SU(6)_W \times$
VMD approach does not necessarily demand violation of Hara's theorem, 
it requires totally
negligible weak $NNM$ couplings if Hara's theorem is to be satisfied. 
 Data on hadronic parity violation
indicate that no such suppression of $NNM$ couplings occurs in reality, however.
Thus, if the pattern $(-,+,+,-)$ of WRHD asymmetries 
(ie., especially, the positive sign of the
$\Xi ^0 \rightarrow \Lambda \gamma $ asymmetry) is confirmed, then,
together with the non-negligible size of
weak $NNM$ couplings this would indicate violation of Hara's theorem.

\section{Acknowledgements}

I would like to thank B. Desplanques for bringing the identity 
of Eq.(\ref{eq:sigmagammaident}) to my attention a few years ago, 
and Ya. Azimov for discussions which prompted this attempt to reconcile
the $SU(6)_W \times$ VMD approach with Hara's theorem.


\end{document}